\documentstyle[aps,psfig,twocolumn]{revtex}

\begin{document}

\draft
\preprint{BNL-HET-99/7, TTP 99-20, hep-ph/9904478}

\title{$\alpha^2$ corrections to parapositronium decay}

\author{Andrzej Czarnecki\thanks{
e-mail:  czar@bnl.gov}}
\address{Physics Department, Brookhaven National Laboratory,\\
Upton, NY 11973}
\author{Kirill Melnikov\thanks{
e-mail:  melnikov@particle.physik.uni-karlsruhe.de}}
\address{Institut f\"{u}r Theoretische Teilchenphysik,\\
Universit\"{a}t Karlsruhe,
D--76128 Karlsruhe, Germany}
\author{ Alexander Yelkhovsky\thanks{
e-mail:  yelkhovsky@inp.nsk.su }
}
\address{Budker Institute for Nuclear Physics,
\\
Novosibirsk, 630090, Russia}
\maketitle
  
\begin{abstract}  
Two-loop QED corrections to the decay rate of parapositronium (p-Ps)
into two photons are presented.  We find $\Gamma(\mbox{p-Ps}\to
\gamma\gamma) = 7989.64(2)\,\mu {\rm s}^{-1}$.  The non-logarithmic
${\cal O}(\alpha^2)$ corrections turn out to be small, contrary to
some earlier estimates.  We speculate that the so-called
``orthopositronium lifetime puzzle'' will not likely be solved by
large QED corrections.
\end{abstract}


Positronium (Ps), the simplest known atom, is an ideal system to test
quantum electrodynamics (QED) of bound states.  The spectrum and
lifetimes of Ps states are, at least in principle, calculable within
QED with very high accuracy.  Hadronic effects, which in other atoms
limit the attainable theoretical precision, are suppressed by the
small ratio of electron and hadron masses.

The lifetimes of the triplet and singlet ground states
(respectively orthopositronium and parapositronium) have been subjected
to very precise theoretical and experimental studies.  Theoretical
predictions for  parapositronium (p-Ps) and orthopositronium (o-Ps) 
decay rates into 2 and 3 photons  respectively, can be expressed as
expansions in the fine structure constant $\alpha$:
\begin{eqnarray}
\Gamma_{\rm p-Ps}^{\rm theory} &=& 
\Gamma^{(0)}_{p}
\left[ 1-\left(5-{\pi^2\over 4}\right) {\alpha\over \pi}
+2\alpha^2 \ln {1\over \alpha} 
\right.
\nonumber \\ && \left.
+ B_p \left( {\alpha\over\pi}\right)^2
- {3\alpha^3 \over 2\pi} \ln^2 {1\over \alpha} 
+\ldots\right],
\label{eq:pps}
\\
\Gamma_{\rm o-Ps}^{\rm theory} &=& 
\Gamma^{(0)}_{o}
\left[ 1-10.286\,606(10){\alpha\over \pi}
-{\alpha^2\over 3}\ln {1\over \alpha} 
\right.
\nonumber \\ && \left.
+ B_o \left( {\alpha\over\pi}\right)^2
- {3\alpha^3 \over 2\pi} \ln^2 {1\over \alpha} 
+\ldots\right],
\label{eq:ops}
\end{eqnarray}
where 
\begin{eqnarray}
\Gamma^{(0)}_{p}={m\alpha^5\over 2},\qquad
\Gamma^{(0)}_{o}={2(\pi^2-9)m\alpha^6\over 9\pi},
\label{eq:born}
\end{eqnarray}
are the lowest order decay widths of the p-Ps and o-Ps, respectively,
and the ellipses in eqs.~(\ref{eq:pps},\ref{eq:ops}) denote unknown
higher order terms which we will neglect in our analysis.  Corrections
of ${\cal O}(\alpha)$ were calculated in \cite{Harris57} for p-Ps.
For o-Ps the most accurate result was obtained in \cite{Adkins96},
where references to earlier works can be found.  The logarithmic
two-loop correction was found in \cite{Caswell:1979vz} for o-Ps and in
\cite{Khriplovich:1990eh} for p-Ps.  The leading logarithmic
correction at three loops was computed in \cite{DL}.
Some partial results on the ${\cal O}(\alpha ^2)$
corrections for both p-Ps and o-Ps  can  be found in
\cite{Adkins96,Burichenko:1992mk,Khriplovich:1994nq,%
Labelle:1994tq,Martynenko:1995vs}, but complete values of $B_{p,o}$
have not been obtained so far. 

Using eqs.~(\ref{eq:pps},\ref{eq:ops},\ref{eq:born}), 
one obtains the following  theoretical predictions for the lifetimes 
\begin{eqnarray}
\Gamma_{\rm p-Ps}^{\rm theory} &=& 7989.42\,\mu {\rm s}^{-1} 
  + \Gamma^{(0)}_p B_{p} \left( {\alpha\over\pi}\right)^2,
\label{eq:ptheor}
\\
\Gamma_{\rm o-Ps}^{\rm theory} &=& 7.0382\, \mu {\rm s}^{-1} 
  + \Gamma^{(0)}_o B_{o} \left( {\alpha\over\pi}\right)^2.
\label{eq:otheor}
\end{eqnarray}

How do these predictions compare with experiments?  For p-Ps the most
recent result \cite{AlRam94},
\begin{eqnarray}
\Gamma_{\rm p-Ps}^{\rm exp} = 7990.9(1.7) \, \mu {\rm s}^{-1},
\label{eq:pexp}
\end{eqnarray}
is in good agreement with (\ref{eq:ptheor}) if  $B_{p}$ is not
too large.  

For orthopositronium the situation is not clear.  Precise experiments of
the Ann Arbor group \cite{Westbrook89,Nico:1990gi} found
\begin{eqnarray}
\Gamma_{\rm o-Ps}^{\rm exp} (\mbox{gas measurement}) &=& 7.0514(14)\,
\mu {\rm s}^{-1}, 
\nonumber \\
\Gamma_{\rm o-Ps}^{\rm exp} (\mbox{vacuum measurement}) &=& 7.0482(16)\,
\mu {\rm s}^{-1} ,
\label{eq:ann}
\end{eqnarray}
which, for $B_{o}=0$, differ from (\ref{eq:otheor}) by
$9.4\sigma$ and $6.3\sigma$ respectively.  This apparent disagreement
of experiment with theory has been known as the ``orthopositronium
lifetime puzzle.''  It should, however, be noted, that a more recent
Tokyo result \cite{Asai:1995re}, 
\begin{eqnarray}
\Gamma_{\rm o-Ps}^{\rm exp} (\mbox{SiO$_2$ measurement}) &=& 7.0398(29)\,
\mu {\rm s}^{-1} ,
\label{eq:tok}
\end{eqnarray}
agrees with the theory prediction if  $B_{o}$ is not
too large.   Both Tokyo and Ann Arbor groups
are working to improve their results.  

Should future experimental efforts confirm the Ann Arbor results
(\ref{eq:ann}), in disagreement
with the QED prediction (\ref{eq:otheor}), the orthopositronium
lifetime puzzle could be solved if $B_o$ turns out to be
unusually large, e.g.~$\sim 250$ for the vacuum 
measurement.  Alternatively, one might speculate that
some ``New Physics'' effects such as o-Ps decays involving axions,
millicharged particles, etc., cause the excess of the measured decay
rate over the QED predictions.  Some of those exotic scenarios seem to
have already been
excluded by dedicated experimental studies. (For a review and
references to original papers see e.g.~\cite{AsaiPhD}.)  

It was anticipated \cite{AlRam94} that a full two-loop calculation
might first be done for p-Ps.  In fact the relative theoretical
simplicity of p-Ps motivated the most recent lifetime measurement
\cite{AlRam94}.  In this paper we present a complete calculation of
$B_p$.  Our result permits the rigorous test of bound state QED
envisioned in \cite{AlRam94}.  We find that the coefficient of the
two-loop non-logarithmic term in (\ref{eq:pps}) is
\begin{eqnarray}
B_p= 5.1(3),
\label{eq:Bp}
\end{eqnarray}
and the theoretical prediction for the p-Ps lifetime becomes
\begin{eqnarray}
\Gamma_{\rm p-Ps}^{\rm theory} &=& 7989.64(2)\,\mu {\rm s}^{-1}.
\end{eqnarray}
Below we briefly discuss some details of our calculation.

The decay width of ${\rm p-Ps} \to 2 \gamma $ can be written as
\begin{eqnarray}
\Gamma &=& \frac {1}{2!~ 4 \pi^2}\sum_{\{\lambda \}} 
\int \frac {{\rm d}^3 k_1}{2\omega _1}
\frac {{\rm d}^3 k_2}{2\omega _2} \delta^4 (P-k_1-k_2) 
\nonumber \\
&&\times
  \left|
  \int \frac {d^3p}{(2\pi)^3} {\rm Tr} \left[ A(\lambda, \mbox{\boldmath$p$})  
 \frac {1+\gamma_0}{2\sqrt{2}}\gamma_5
 \right]~\phi(\mbox{\boldmath$p$})
  \right|^2,
\label{width}
\end{eqnarray}
where $P$ is the four-momentum of the p-Ps,
$\phi(\mbox{\boldmath$p$})$ is its wave function, and
$A(\lambda,\mbox{\boldmath$p$})$ 
is the annihilation amplitude of an $e^+ e^-$ pair into a pair of photons
with polarization $\{ \lambda \}$.

In the non-covariant perturbation theory the on-shell
amplitude of the process $e^+e^- \to \gamma\gamma$ reads
\begin{eqnarray}
A &=&
 {8 \pi \alpha E_p\over E^2_{p-k}(E_p+m)}
    \mbox{$\Lambda_-({\mbox{\boldmath$p$}})$}
\vec{\alpha}\cdot \mbox{\boldmath$e$}_2
    \left[ \vec{\alpha}\cdot (\mbox{\boldmath$p$}-\mbox{\boldmath$k$})
 + \beta m \right] 
\nonumber \\ && \times
    \vec{\alpha}\cdot\mbox{\boldmath$e$}_1
         \mbox{$\Lambda_+(\mbox{\boldmath$p$})$} 
 + (\mbox{\boldmath$e$}_2 \leftrightarrow \mbox{\boldmath$e$}_1,
    \mbox{\boldmath$k$} \leftrightarrow -\mbox{\boldmath$k$}).
\label{A}
\end{eqnarray}
Here $E_p=\sqrt{m^2+\mbox{\boldmath$p$}^2}$; $\mbox{\boldmath$p$}$ and
$\mbox{\boldmath$k$}$ are electron and photon 
three-momenta in the p-Ps rest frame;
and $\Lambda_{\pm}(\mbox{\boldmath$p$})$  are the projectors on the 
positive and negative energy states, 
\begin{equation}
\Lambda_{\pm} = \frac{1}{2} \left( 1 \pm
                \frac{\vec{\alpha}\mbox{\boldmath$p$} + \beta m}{E_p}
                \right).
\end{equation}
To leading order one can neglect the small momenta $p \sim m\alpha$
compared to $m$ and $|\mbox{\boldmath$k$}| \sim m$. One finds the
following leading order amplitude
\begin{equation}
A_{\rm LO} = -\frac {4\pi\alpha}{m^2}  
(\vec{\alpha} \mbox{\boldmath$e$}_2)(\vec {\alpha}
\mbox{\boldmath$k$})(\vec{\alpha} \mbox{\boldmath$e$}_1)  ,
\end{equation}
and the lowest order decay width
\begin{equation}
\Gamma_{\rm LO} = \frac {4 \pi \alpha^2}{m^2} \left| \psi(0)\right|^2 = 
\frac {m\alpha^5}{2}.
\label{LO}
\end{equation}

Higher order corrections to $\Gamma_{\rm LO}$ will be calculated using
non-relativistic QED (NRQED) \cite{Caswell:1986ui} with dimensional
regularization \cite{Pineda:1997bj}.  
We divide up the corrections into three parts:
\begin{eqnarray}
B_{p} &=& B_{p}^{\rm squared} + B_{p}^{\rm hard} + B_{p}^{\rm soft},
\end{eqnarray}
where $B_{p}^{\rm squared}$ is the contribution of the one-loop
amplitude squared and $B_{p}^{\rm {hard,~soft}}$ are the hard and soft
contributions.  The hard corrections arise as contributions of virtual
photon  momenta $k \sim m$.  Their effects are described by adding
operators containing $\delta(\mbox{\boldmath$r$})$ to the
non-relativistic Hamiltonian. 
The technical challenge is to compute the Wilson coefficients of those
operators.  Fortunately those coefficients can be obtained using any
convenient external states.  In particular one can compute them for
the electron and positron at rest.  It is important to employ
dimensional regularization, so that one avoids the necessity of
subtracting the Coulomb singularities from box graphs.  On the other
hand, the soft contributions come from the region of virtual photon
momenta of the order of $k \sim m\alpha$ and are sensitive to bound
state dynamics.  The actual calculation is briefly described
below.

The square of the one-loop amplitude is easily obtained from the
one-loop result:
\begin{eqnarray}
B_p^{\rm squared}=\left({5\over 2}-{\pi^2\over 8}\right)^2 \simeq 1.6035.
\label{eq:bpsquared}
\end{eqnarray}

\begin{figure}[h]
\hspace*{-10mm}
\begin{minipage}{10.cm}
\[
\mbox{
\begin{tabular}{ccc}
\psfig{figure=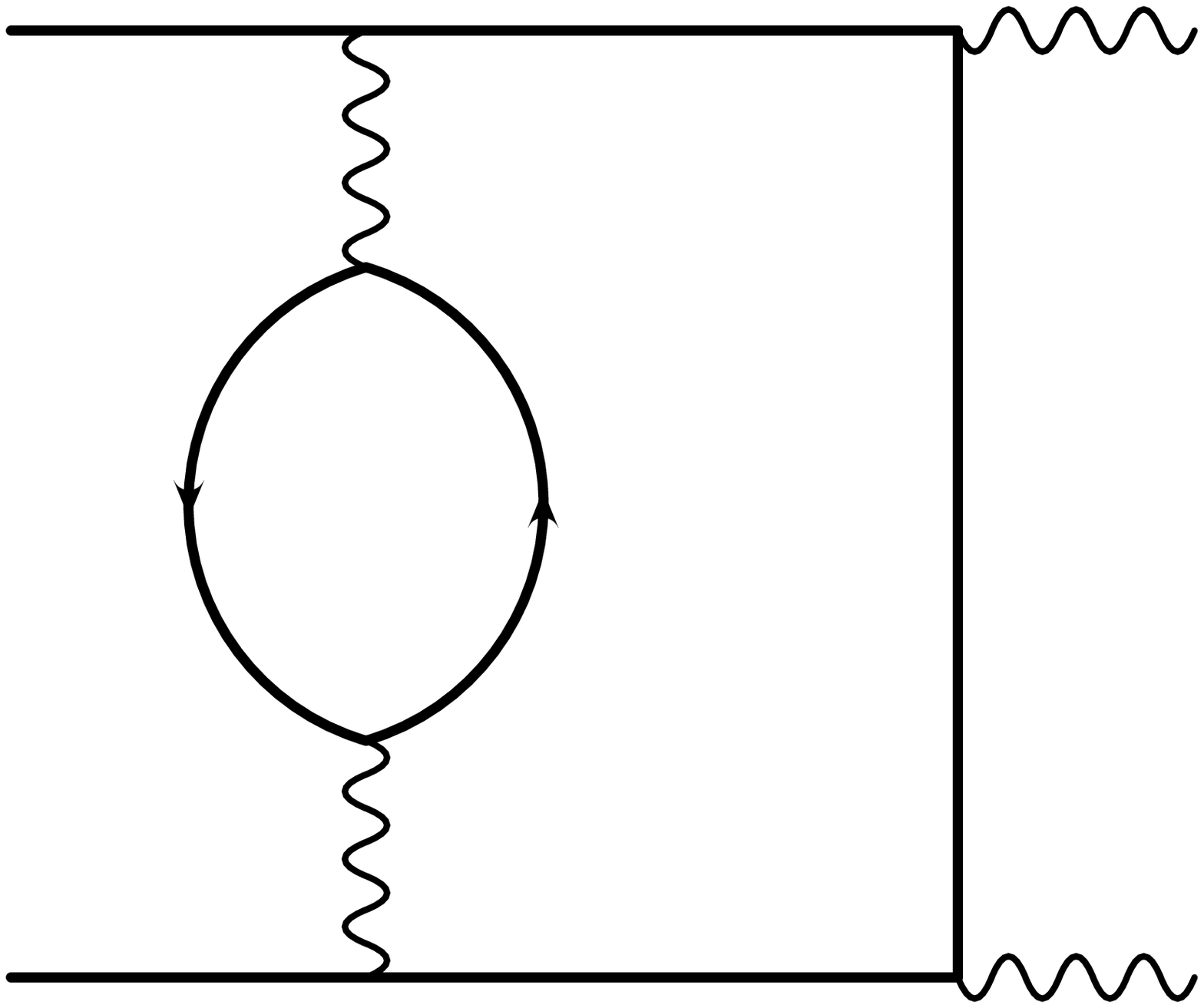,width=23mm}
&\hspace*{2mm}
\psfig{figure=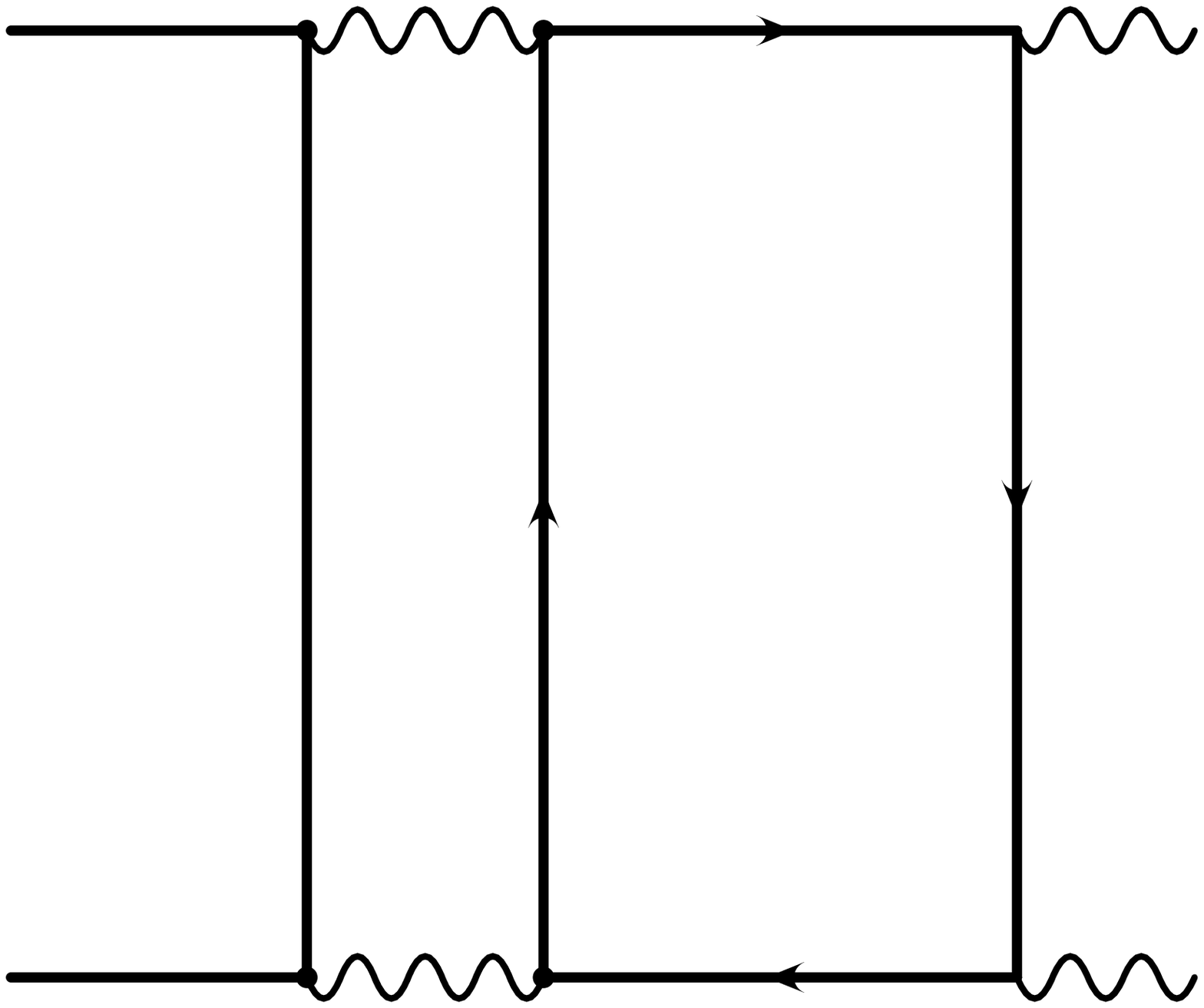,width=23mm}
&\hspace*{2mm}
\psfig{figure=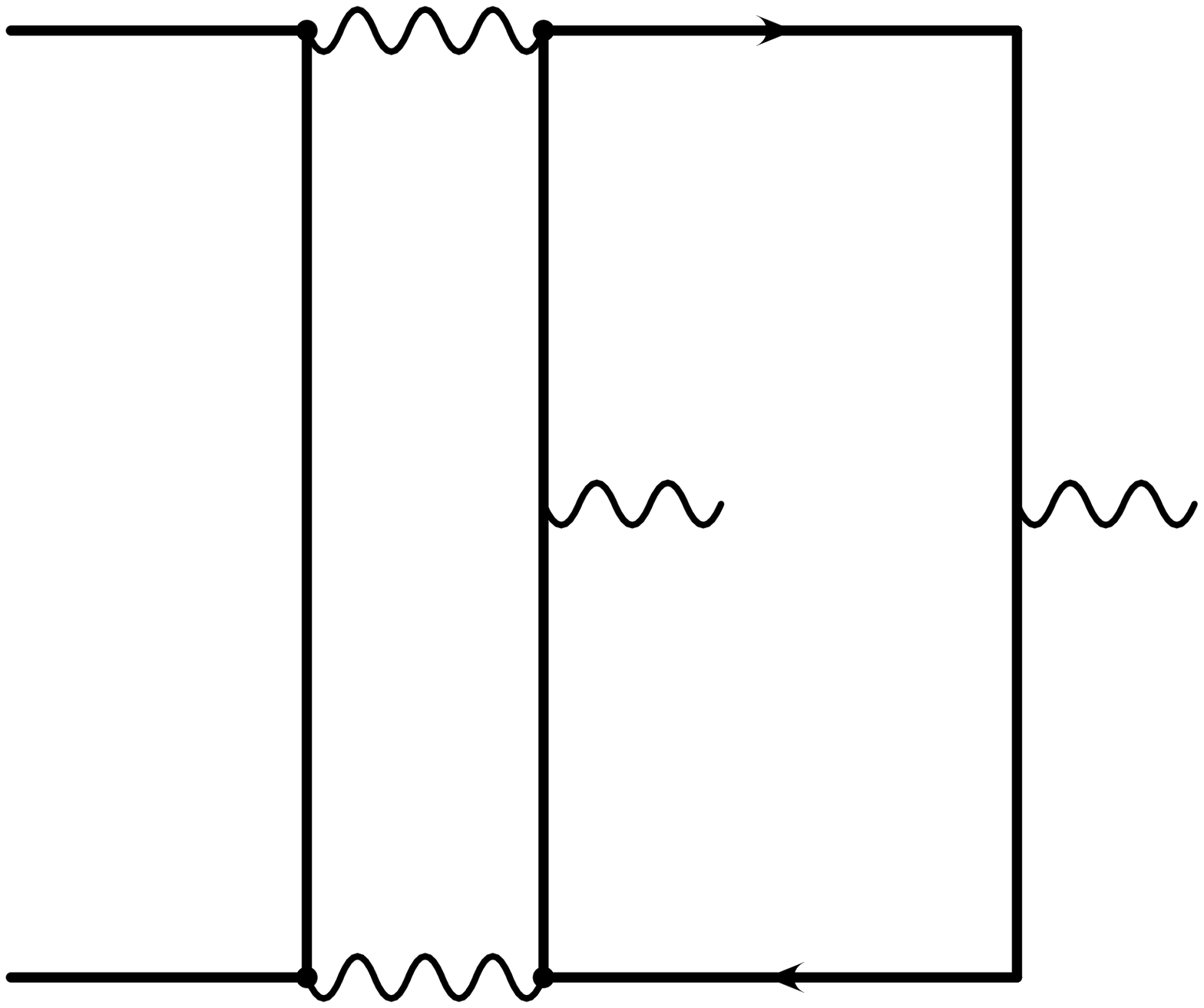,width=23mm}
\\[1mm]
 (a) & (b) & (c)
\\[2mm]
\psfig{figure=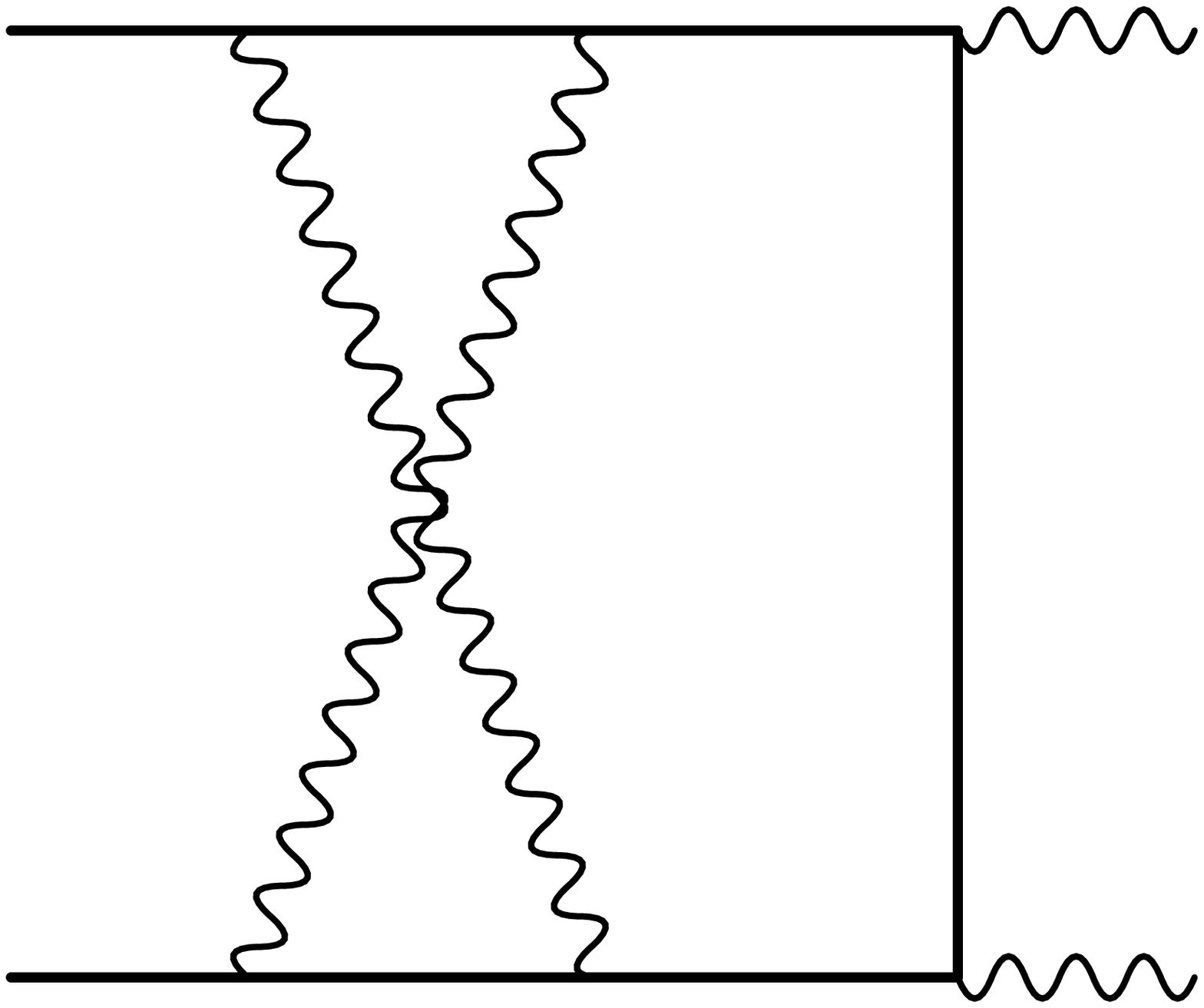,width=23mm}
&\hspace*{2mm}
\psfig{figure=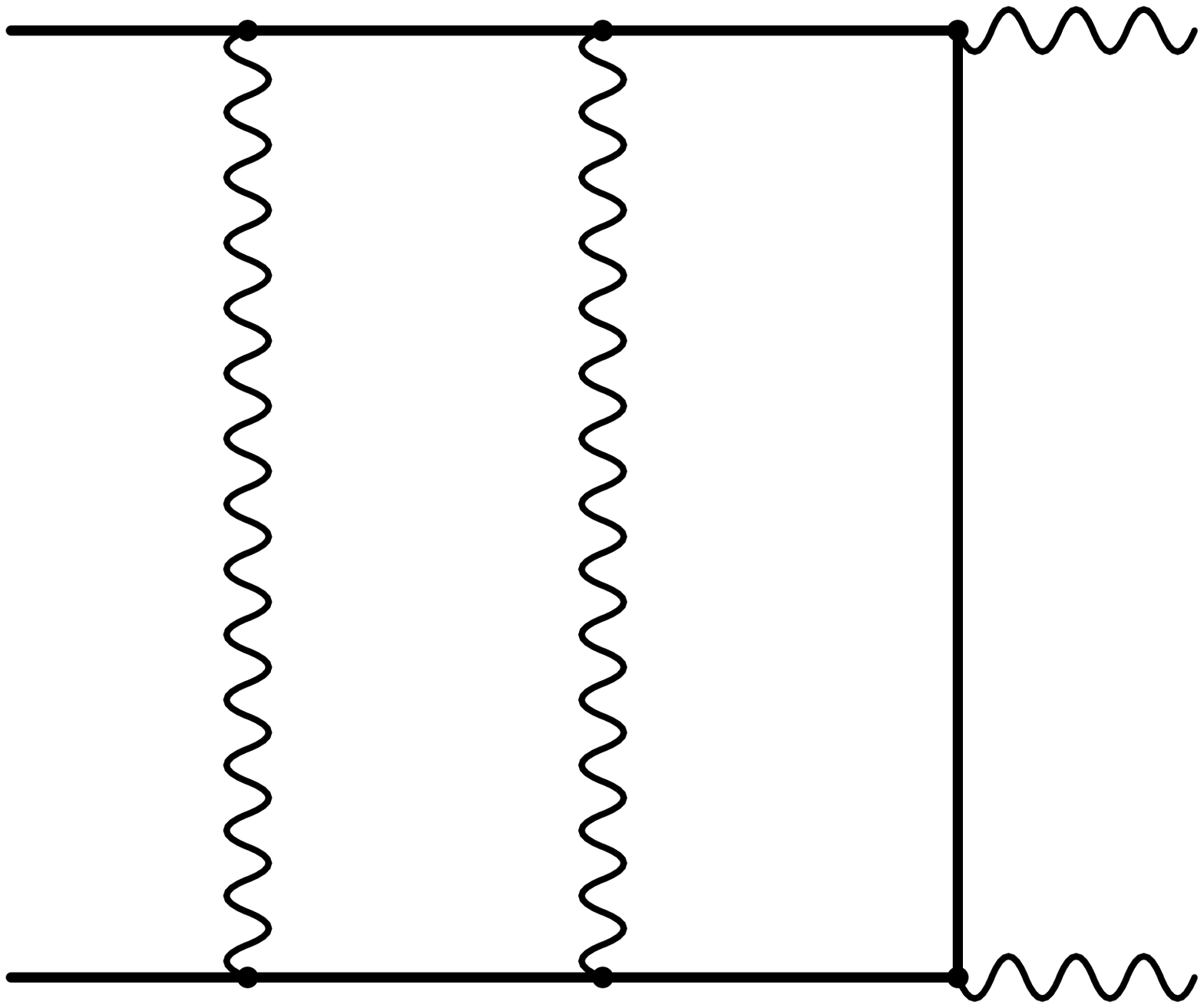,width=23mm}
&\hspace*{2mm}
\psfig{figure=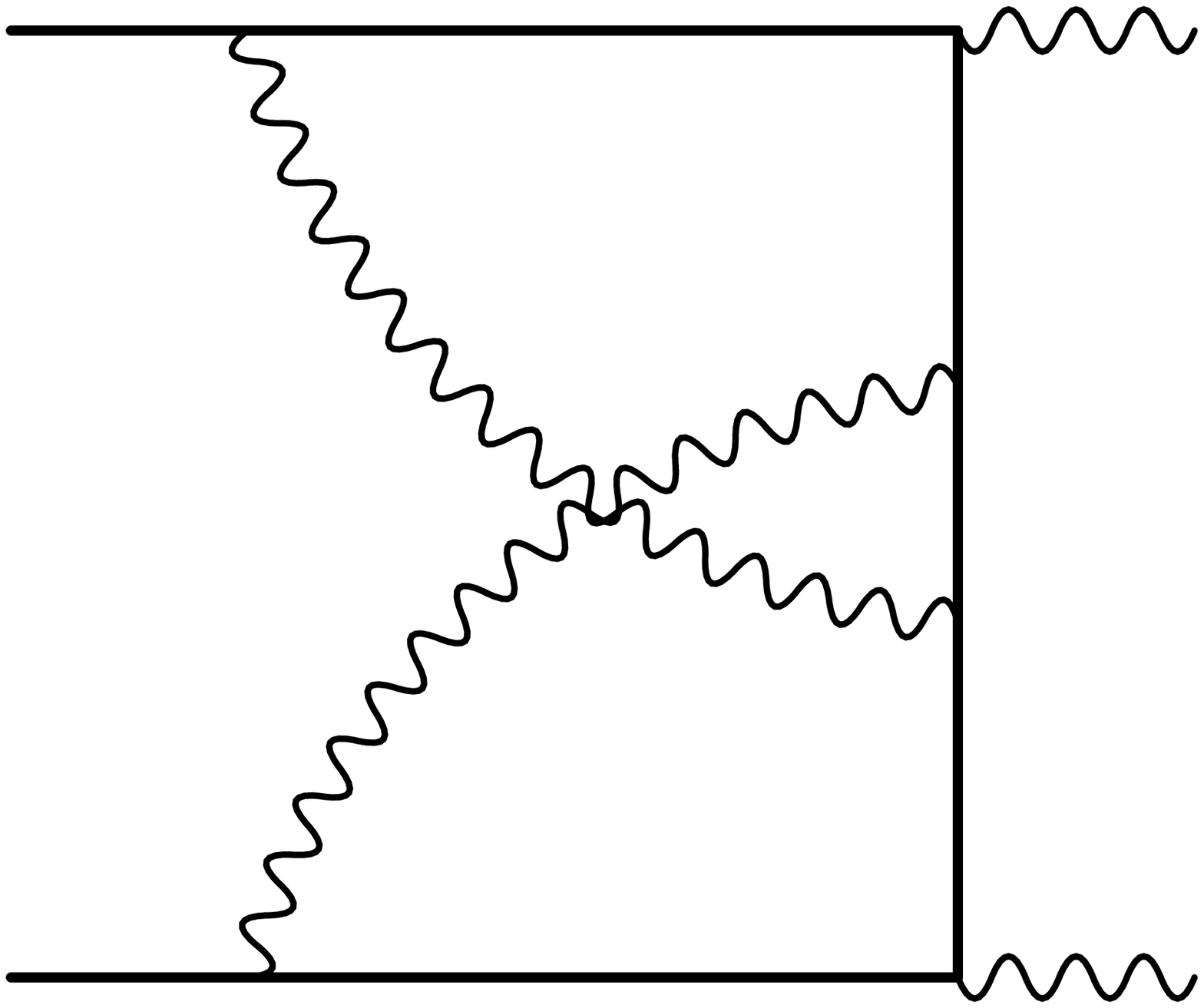,width=23mm}
\\[1mm]
  (d) & (e) & (f)
\\[1mm]
\end{tabular}
}
\]
\end{minipage}
\caption{Examples of two-loop hard corrections to p-Ps decay into two
photons.} 
\label{fig:2loop}
\end{figure}

$B_p^{\rm hard}$ consists of three types of contributions:
vacuum polarization insertions in the photon propagators,
light-by-light scattering diagrams, and two-photon corrections to the
annihilation amplitude,
\begin{eqnarray}
B_p^{\rm hard}=B_p^{\rm hard}(\mbox{VP}) 
+ B_p^{\rm hard}(\mbox{LL})
+ B_p^{\rm hard}(\gamma\gamma).
\end{eqnarray}
Vacuum polarization insertions into the one-loop
graphs (an example is shown in Fig.~\ref{fig:2loop}(a)) 
were computed in \cite{Burichenko:1995as,AdkShif},
\begin{equation}
B_p^{\rm hard}(\mbox{VP})= 0.4468(3).
\label{VP}
\end{equation}
Light-by-light scattering contributions  (for examples see
Fig.~\ref{fig:2loop}(b,c)) 
are more difficult to compute because of their imaginary parts
which make numerical integration unstable.  We computed them
analytically, by formally assigning a large mass $M$ to the internal
fermions  and expanding in $x = m/M$.   The resulting series 
converge so well that several terms  are sufficient  to find 
the result at $x=1$:
\begin{equation}
B_p^{\rm hard}(\mbox{LL})= 1.28(13).
\label{lbl}
\end{equation}
The most difficult class of effects comes from the two-photon
corrections, examples of which are shown in Fig.~\ref{fig:2loop}(d-f).
We proceed in the following way: combine propagators using Feynman
parameters; perform momentum integrations analytically; extract
ultraviolet (UV) and infrared (IR) divergent pieces; integrate
numerically over typically 5 (in some cases 6) Feynman parameters in
the finite expressions.

Extraction of UV divergences is relatively simple -- they
show up, roughly speaking, as singularities in the overall factors
rather than as divergent integrals over Feynman parameters.

Calculation of IR divergent diagrams is more demanding.  For each such
diagram we add and subtract a simpler diagram so that the sum is IR
finite, and the subtraction can be calculated analytically.  As an
example let us consider the diagram shown in
Fig.~\ref{fig:2loop}(d). The IR singularity in this diagram appears
when momenta of the virtual photons are small.  To suppress the
contribution of this region we rewrite the propagator of the
$t$-channel electron by subtracting its value 
and a suitable number of derivatives
when both virtual photon
momenta are zero.  This difference leads to an infrared finite
expression which can be computed numerically.  To compensate this
subtraction we must add the same diagram with the $t$-channel
propagator replaced by a constant, $1/(2m^2)$.  Such diagrams were
studied in \cite{threshold} and can be computed analytically.

Similar tricks are used to compute all other IR divergent diagrams,
although the subtraction procedure is more tedious in case of diagrams
with stronger singularities, like the planar box in
Fig.~\ref{fig:2loop}(e).

Adding all two-photon diagrams and summing up 
numerical errors of individual diagrams in quadrature we find
\begin{eqnarray}
B_p^{\rm hard}(\gamma\gamma) = - {\pi^2\over 2\epsilon} 
+2\pi^2\ln m
-42.23(27).
\label{gamgam}
\end{eqnarray}
The logarithm of the dimensionfull parameter $m$ 
arises from the expansion of the overall factor $m^{-2\epsilon}$
and vanishes in all physically meaningfull expressions.

The sum of eqs.~(\ref{VP}), (\ref{lbl}), and (\ref{gamgam}) gives
the total hard correction 
\begin{eqnarray}
B_p^{\rm hard} = - {\pi^2\over 2\epsilon} +2\pi^2\ln m
-40.46(30).
\label{hard}
\end{eqnarray}

To calculate the soft scale contributions, one should account 
for the relativistic corrections 
to the annihilation amplitude (AA)
$e^+e^- \to \gamma \gamma$ and
to the positronium wave function (WF):
\begin{eqnarray}
B_p^{\rm soft} = B_p^{\rm soft}(AA) +  B_p^{\rm soft}(WF).
\end{eqnarray}

For the annihilation amplitude correction, 
one expands the on-shell amplitude (\ref{A}) 
to relative order ${\cal O}(p^2/m^2)$.  Although the
resulting integral is linearly divergent, using dimensional
regularization one finds a finite result 
(see \cite{Czarnecki:1999mw} for a discussion of this effect):
\begin{eqnarray}
B_p^{\rm soft}(AA) = {\pi^2\over 3}.
\label{ga}
\end{eqnarray}

Relativistic corrections to the positronium wave function can be
computed using the Breit Hamiltonian. Since we regularize all
divergences dimensionally, we need the Breit Hamiltonian in
$d$-dimensions derived in \cite{Czarnecki:1999mw}.  Its projection on
the $S$-states can be found in eq.~(39) of that paper.  Performing
calculations similar to those described after that equation, we find
the wave function correction to the decay rate:
\begin{eqnarray}
B_p^{\rm soft}(WF)
+2\pi^2\ln{1\over \alpha}
= {\pi^2\over 2\epsilon} + 2 \pi^2 \ln{1 \over m\alpha} + \frac {33\pi^2}{8} ,
\label{gb}
\end{eqnarray} 
where on the LHS we have separated the logarithm, to be consistent
with the division of corrections introduced in eq.~(\ref{eq:pps}). 

The sum of the corrections to the 
annihilation amplitude (\ref{ga})
and to the wave function (\ref{gb}) gives the final result for the
soft contributions,
\begin{equation}
B_p^{\rm soft} = 
 \frac {\pi^2}{2\epsilon} - 2\pi^2 \ln m + \frac{107\pi^2}{24} . 
\label{soft}
\end{equation} 
We note that this partial result cannot be directly compared to the
soft corrections found in a previous study 
\cite{Khriplovich:1994nq} since a 
different regularization scheme was employed there.

The final result in eq.~(\ref{eq:Bp}), $B_p = 5.1(3)$, is obtained
as a sum of the square of the one-loop corrections 
(\ref{eq:bpsquared}), and the
genuine two-loop hard (\ref{hard}) and soft (\ref{soft}) 
contributions.  We can now present a theoretical prediction for the
2-photon width of parapositronium with the 2-loop accuracy:
\begin{eqnarray}
\Gamma_{\rm p-Ps}^{\rm theory} &=& {m\alpha^5\over 2}
\left[ 1-\left(5-{\pi^2\over 4}\right) {\alpha\over \pi}
+2\alpha^2 \ln {1\over \alpha} 
\right.
\nonumber \\ && \left.
+5.1(3)\left( {\alpha\over\pi}\right)^2
- {3\alpha^3 \over 2\pi} \ln^2 {1\over \alpha} 
\right] 
\nonumber \\
&=& 7989.64(2)\,\mu {\rm s}^{-1}.
\label{th}
\end{eqnarray}
Our final result (\ref{th})   agrees 
well within $1\,\sigma$ with the most recent experimental 
result, eq.~(\ref{eq:pexp}).  

p-Ps can also decay into 4 or more photons.  Those effects increase
the p-Ps width by about $0.01 \,\mu {\rm s}^{-1}$
(see \cite{Lepage:1983yy} and references therein).

The coefficient of the non-logarithmic $(\alpha/\pi)^2$
term in eq.~(\ref{th}) is rather small, due to an almost complete
cancelation between the soft and hard corrections.  
As we have already mentioned, only the sum of the two is regularization scheme
independent and hence unambiguous. For this reason,  
it is likely that the cancellation between soft and hard pieces 
is not accidental.  Scheme and gauge
independent corrections, for example eqs.~(\ref{eq:bpsquared},
\ref{VP}, \ref{lbl}), seem to indicate that
``natural scale'' of the ${\cal O}(\alpha^2)$ corrections is $[\rm
several~units] \times (\alpha/\pi)^2$. 

The result of our calculation, $B_p=5.1(3)$, is much
smaller than the estimate $40\pm 20$ given in \cite{Khriplovich:1994nq}.
This discrepancy can be traced back to the discussion after 
eq.~(26) in the first paper of \cite{Khriplovich:1994nq}.
It seems that the impact of short-distance (hard) corrections
was underestimated there. As the division of contributions 
into soft and hard pieces is regularization scheme dependent,
it is clearly dangerous to draw conclusions about the complete
result on the basis of only one of those parts.

Having for the first time a complete two-loop correction to a QED
bound state decay, one is tempted to speculate about the size of such
corrections to the orthopositronium lifetime.  Although nothing can be
said rigorously, one could argue that most known second order effects
have similar order of magnitude for o-Ps and p-Ps.  A possible
enhancement may be due to the larger (by about a factor of 3) number
of Feynman diagrams contributing to o-Ps decay compared to p-Ps decay
(this is already seen in the magnitudes of the ${\cal O}(\alpha)$
corrections).  Unless this enhancement is dramatic for the complete
${\cal O}(\alpha^2)$ corrections to the decay rates, the theoretical
prediction for the o-Ps lifetime will remain distant from the
experimental results in eq.~(\ref{eq:ann}).  It is therefore extremely
important that the 3-photon decay of o-Ps be further studied
experimentally.

We thank William Marciano for carefully reading the manuscript and
helpful remarks.
We are grateful to G.~Adkins, R.~Fell, and J.~Sapirstein for pointing
out an error in our initial evaluation of the light-by-light
contributions (\ref{lbl}). 
This research was supported in part by DOE under grant number
DE-AC02-98CH10886, by BMBF under grant number BMBF-057KA92P, by
Gra\-duier\-ten\-kolleg ``Teil\-chen\-phy\-sik'' at the University of
Karlsruhe, by the Russian Foundation for Basic Research under grant
number 99-02-17135 and by the Russian Ministry of Higher Education.



\end{document}